**Co-opted marginality, a new type of anti-immigrant discourse on social media?**

**Classifying social media messages about immigrants with BERT**

Claire Stravato Emes
Wee Kim Wee School of Communication
Nanyang Technology University
claire001@e.ntu.edu.sg

Anfan Chen
School of Journalism and Communication
The Chinese University of Hong Kong
anfanchen@cuhk.edu.hk

## Abstract

The dominant frame about immigrants in the public sphere is often depicted to consolidate low "common sense" beliefs about immigrants being a social problem and emphasizing the superior status of citizens (Gabrialatos & Baker; 2008; Kim & al. 2011; Qinsaat, 2014). The view that anti-immigrant discourse may not solely be an expression of xenophobia, prejudices or dominant power relationships has rarely been questioned.

The article scrutinizes public discourse in comments of 11 social media community platforms over six months in Singapore. We conducted a qualitative framing analysis to synthesize Singaporeans' viewpoint about immigrants in online comments sections, followed by quantitative textual analysis. Over 300,000 comments were collected and classified via manual coding and machine learning (BERT). The results show that most comments discussing immigrants online do not indicate xenophobia but instead express fears of ingroup members being marginalized.

The study identifies a new type of rhetoric for anti-immigrant discourse, which we name co-opted marginality, whereby a majority/dominant group claims to be marginalized. Ultimately, the findings suggest that anti-immigrant discourse may be attributable to factors other than xenophobia (fear of immigrants), such as identification issues and perceived loss of status.

## Introduction

Social media's socio-technical structure allows the deployment of discursive spaces, which like traditional coffee houses and churches, shelter the emergence of new discourses (Scullion et al., 2013). The Internet has disrupted the linearity principles around frame building (Riles et al., 2017). New voices intervene in the process as citizens engage and upset existing gatekeepers' ability to infuse frames. Commenting on the online communities and news sites have become an essential form of citizen engagement (Amlgren & Olsson, 2016; Noci et al., 2012; Reich, 2011; Rutz et al., 2011; Thurman, 2008). Since the mid-2000s, comment sections operating on mass media outlets or community sites have been depicted as new forms of public spheres that provide communicative spaces and allow sub-counter publics to contest the discursive boundaries of the dominant rhetoric (Asen, 2000). According to Toefl & Piwoni (2015), comment sections in community public spheres are sites that challenge dominant consensus structures and reveal new types of critical discourse. Despite possibly displaying low deliberative quality (Dahlberg, 2011; Freelon, 2015), comments in the online public sphere constitute a collective thinking structure that emphasizes divergence and adds nuance to the dominant rhetoric.

Anti-immigrant rhetoric is generally deplored by the dominant rhetoric of mainstream press and political parties (Alduy, 2017; Kirk & Martin, 2017; Lind, 2019; Reilly, 2016). Remarks attributed to presidential candidate Hillary Clinton in 2016 are symptomatic of the attitude: "You know, to just be grossly 'generalistic,' you could put half of Trump's supporters into what I call the basket of deplorables. Right? They're racist, sexist, homophobic, xenophobic, islamophobic—you name it." (Reilly, 2016).  Lind (2019) argues that anti-racist, liberal, and democratic forces easily associate overt and insistent patriotism with authoritarians.

However, we argue that analyzing anti-immigrant discourse within the online public sphere may reveal other viewpoints about social reality beyond the argumentative themes belonging to authoritarian and racist grammar. This paper focuses on deconstructing the discursive elements about anti-immigrant discourse online in the specific context of online comments on public forums. The objective is to analyze whether commenting spaces reflect a more nuanced and multifaceted discourse about immigrants than the one often reflected in the dominant consensus shaped by political leaders and the press.

## Literature Review

**Comment section: a sub-counter public sphere to observe collective interpretations**

The comment sections encourage counter public activity (Freelon, 2015; Toepfl & Piwoni, 2015, 2018). Teopfl and Piwoni (2015) identified that three-quarters of all online articles' comments on the anti-euro movement in the German online public sphere sharply opposed the mainstream editorial lines of Germany's unanimously pro-common-currency media. By commenting, people challenge the orientations of stories, leading to re-interpretation and the emergence of spontaneous and unintentional frames (Riles et al., 2017); it also allows individuals to change the original value of the news discussed online (Ziegele et al., 2013). Whatever the user's motivations, their comments mix individual and discussion-dependent factors that might match, accentuate, or denigrate the news factors in the original posts (Moussaïd, Brighton & Gaissmaier, 2015).

Participation in communities online also bolsters citizens' desire to participate in countercultural discussions and can improve individuals' sense of collective efficacy (Halpern et al., 2017). They develop a collective agency that reinforces their desire to participate and build knowledge together. Participants can know what others think (Huckfelt, 2014), share political content on social media, and learn about their community's essential social and civic matters (Bode et al., 2014).

From the perspective of observers, the comment section provides an opportunity to give a clear account of people's reality. Moments of dispute and collective interactions online reveal participants as agents who can innovatively perform outside the control of any gatekeepers. Following Boltanski's (2006) principles of the pragmatic sociology of critique, the comments section may allow actors to prioritize what to think, say, and do in their situated contexts and their day-to-day interactions. While commenting, actors expose conditions of their existence and of their criticisms towards the reality they encounter. We assume that interactions online provide a window with no abstractions about personal takes on realities and collectives. Commenting sections online can reveal how ordinary people identify paradoxes they face and create collectives within their community. Echoing Boltanski and Claverie (2007), the online community forum's context allows reproducing a "social world as the scene of a trial" (p. 395).

Ultimately, comment sections online allow for detecting the elements and processes associated with a collectively shared discourse that may challenge the dominant consensus. Before considering whether comment sections may shelter new types of anti-anti-immigrant discourse, it is fitting to scrutinize the literature in general on the anti-immigrant discourse on social media.

**Discourse about immigrants on social media**

Over the years, discourse about immigrants on social media has been scrutinized under the lens of populism (Ciani & Parent, 2016; Ernst et al., 2018; Wirz & al, 2018), political extremism (Atton, 2006; Ernst & Sina, 2019), racism (Bliuc et al. 2018; Klein et al., 2018) and authoritarianism (Erjavec et al., 2012; Fuchs, 2018). Traditionally racist and discriminatory discourse is seen as explicit, aiming at strengthening convictions around ingroup memberships by framing prejudice as a natural response (Caiani & Parenti, 2016; Weaver, 2013).

However, it appears that the strategies used by actors to convey such nativist and racist discourse online have morphed into an increasingly sophisticated and unassuming discourse (Alduy, 2017; McNamee et al., 2010; Tynes et al., 2012). Lately, approaches are more subtle; promoters of prejudicial thinking tend to trivialize, denigrate, and even deny its existence (Cresswell et al., 2014; Durrheim et al., 2014; Goodman & Rowe, 2014).

A salient and seemingly undetected form of anti-immigrant discourse worldwide is articulated by individuals who profess the grammar of victimhood and claim that immigrants obtain advantages over the citizens' needs. In America (Server, 2019), Scandinavia (Herkman, 2016; Laaksone et al., 2020), Western Europe (Gattinara, 2018), and Singapore (Chin, 2017; Teng, 2019), citizens argue refugees or immigrants are given more legitimacy and resources, a claim summarized under the slogan, "citizens are second class." Such marginalization frame appears perplexing because it contradicts the social identity theory in which in-groups (citizens) benefit from high-status in their homeland (Walzer, 1983). Supporters of the marginalization frame adopt a discourse inconsistent with their group's relatively higher-status position in society. Conventionally, outsiders that challenge the established status hierarchy get penalized for any status violations (Betz, 2017). Such responses reveal a willingness of the dominant group members to assert and comfort dominance, not acknowledging it is faltering. Instead of reinforcing and asserting dominance, co-opted marginality supporters advocate a position of underdogs despite unclear evidence of actual discrimination (Earle & Hodson, 2020).

These claims, which we refer to as co-opted marginality, focus on citizens' inability (ingroup members) to compete and assert their dominance; they reflect citizens' feelings of insecurity about their positions. Contrary to anti-immigrant rhetoric, the stance does not assume immigrants' (outsiders) assumed inferiority.

Research on cyber-racism and discriminatory discourse online still questions the role of social media in mobilizing isolated individuals and enabling ideological rapprochement of supporters around reactionary ideas, such as anti-migrant sentiment. A research gap persists in understanding why and how individuals who do not identify as extremists nor racists communicate populist views or rally behind anti-immigrant discourse on social media (Bliuc et al., 2018). Within the context of communication on social media, the marginalization frame may offer an interesting explorative path for addressing why isolated individuals, who are not formally affiliated to extremist groups or viewpoints, profess reactionary claims. The co-opted marginality claim may be compelling to ingroup members who may not be inclined to prejudicial thinking but may feel marginalized by immigrants' presence (Ribuffo, 2017).

## The Singaporean Context

We explore anti-immigrant discourse in Singaporeans online public sphere. Singapore has recently experienced a surge in anti-migrant feelings. The city-state has built one of the strongest economies in the world by importing an immigrant workforce. Immigration in Singapore is necessary to compensate for the lack of natural resources (Lim, 2019). The population grew by 40% in the last 20 years, well ahead of significant metropolises like London or New York (Worldbank, 2020), opening its border to economic migrants. These include "Foreign Talents," a term that designates ambitious, high-skilled foreign individuals who provide intellectual capital to fuel the city's growth. The general election in June 2020 propelled the issue at the forefront of the public debate, as the opposition openly articulated the idea: "No to 10 million population—Don't allow more immigrants to come in to compete for our jobs." (Straits Times, 2020).

The local press generally rebukes the expression of anti-immigration and nativist sentiments, sympathetic to the pro-immigrant government. In Singapore, the gatekeepers have assimilated negative discourse about immigrants as a malignant discourse and judge it as an irresponsible and dangerous outburst of the putrid citizenry (Chin, 2017; Straits Time, 2014). The discourse has been condemned even more given strong multicultural principles in Singapore. Singaporeans are subjected to the "OB markers" (out of bound markers), whereby individuals tacitly know what is taboo and what should not be discussed publicly (Hallin & Mancini, 2011). Race-related issues are avoided in public discourse; a young woman received a 10-month jail sentence in 2016 for posting on a website called "the real Singapore" articles that were intended to "incite ethnic hatred" (Ramzy, 2016). Still, as xenophobic rhetoric is reported to increase in Singapore on social media (Washington Post, 2020), it is fitting to question the type of discourse about immigrants that emerges in the online public sphere.

Singapore does not officially censor the Internet (Seng, 2008); critical websites and foreign media sites are freely accessible. However, restrictions are in place, such as registration fees for the mandatory licensing of online political sites, which give the Infocomm Media Development Authority (IMDA) the power to request political sites to remove content deemed dangerous to Singapore's political stability. Since 2010, Singaporeans have had access to a wide choice of news community sites apart from government-backed mainstream broadcasting and news media (Channel News Asia, Straits Times), many independently financed news websites have emerged, some of them adopting at times an openly-critical stance towards government policies (State Time Review, Online Citizen Online). In February 2019, The State Time Review (STR) and The Online Citizen (TOC) were investigated for criminal defamation over the publication of an article over a local MP in a 1MB criminal Malaysian case. The TOC was de-gazetted as a political association. The turn of events prompted the Ministry of Law to emphasize the need to introduce legislation, POFMA, to stop fake news (Varswani, 2019).

Considering the theoretical arguments discussed earlier, we expect that critical discourse is homogeneously distributed in the online public sphere's comment sections regardless of the online site's position towards the dominant consensus structure.

RQ1: What type of rhetoric about immigrants is salient in Singapore's online community public forum comments section?

RQ2: What types of discourse attract the most reactions when commenting about immigrants in Singapore's online community public forum comments section?

RQ3: Are the different anti-immigrant discourses present and homogeneously distributed in Singapore's online community public forum comments section?

Finally, we identified that the co-opted marginality frame, used when discussing immigration issues, upsets the predictions of social identity theory. For example, when group members discuss immigrant-related matters, Singaporean identification discourse should reflect negative sentiments linked to identification as a Singaporean. We, therefore, test the following hypothesis:

H4 (a): Rhetoric about immigrants is more positively associated with positive than negative sentiments linked to Identification.

H4 (b) & (c): Rhetoric about authorities (b), co-opted marginality (c), are more positively associated with negative sentiments linked to Identification.

## Method

### *Data Collection*

We selected popular online forums with high numbers of followers and likes as a proxy for the Singaporean online public sphere's representativity to answer our questions. The selection included 11 Facebook sites of popular community sites (See Table 1) located on a broad ideological spectrum: one national newspaper backed by the government, Straits Times; one broadcaster, Channel News Asia; five community sites including Stomp, Mothership, All Stuff Singapore, SMRT, MustShareNews, Reddit/Singapore and three alternative news sites: TOC, Independent Sg, and STR.

**Table 1**

11 Singapore Popular Community Sites Included in Our Study

| Community Site | Description | Follower Number (2019) |
|---|---|---|
| *Straits Times* | *The English-language daily broadsheet newspaper based in Singapore and currently owned by Singapore Press Holdings, https://www.straitstimes.com/* | 1.2 m followers |
| *Channel News Asia* | *English-language news channel based in Singapore, https://www.channelnewsasia.com/news/international* | 2.7 m followers |
| *Stomp* | *Social networking and citizen media website stands for Straits Times Online Mobile Print, https://stomp.straitstimes.com/* | 587 k followers |
| *Mothership* | *A Singapore-born Internet media company with the highest local penetration among all digital-only media platforms in Singapore, https://mothership.sg/* | 262 k followers |
| *All Stuff Singapore* | *Singapore's fastest-growing news portal. Self-portrayed as independent and not affiliated to any media organization or political party, https://www.allsingaporestuff.com/* | 398 k followers |
| *SMRT Feedback by The Vigilanteh* | *Independent Facebook-based community site, https://www.facebook.com/smrtsg/* | 431 k followers |

| MustShareNews | *Online news site for the social media generation, covering the most important stories for millennials. https://mustsharenews.com/* | 67 k followers |
|---|---|---|
| *Reddit (Singapore)* | */r/Singapore, Reddit home of the country Singapore., https://www.reddit.com/r/singapore/* | 277 k followers |
| *Independent Sg* | *Independent news platform, https://theindependent.sg/* | 64 k followers |
| *The Online Citizen* | *Singapore's longest-running independent online media platform. charged with criminal defamation over corruption allegation (2019), https://www.onlinecitizenasia.com/* | 105 k followers |
| *State Time Review* | *Anti-establishment website States Times Review shutdown end of 2019, http://www.str-subsea.com/contact/singapore* | 50 K followers |

Comments, along with metadata on shares, likes, comment counts, and post IDs, were collected from June 1st, 2018, to December 1st, 2018, before the enactment by the Minister of Law of the bill to protect society from online falsehoods and malicious actors which makes posting "deliberate online falsehoods" a criminal offense. A social media data extracting tool, Netvizz (http://www.up2.fr/index.php?n=Main.Netvizz) (Rieder, 2013; Spry, 2019), was used to retrieve comments on Facebook. Discussions threads in Reddit were scraped through an Application Programming Interface using the Python package Praw. Through this process, 1,598 posts and 424,713 comments were yielded, 386,922 comments were retained after removing the duplicate comments and null comments (The Entire Corpus in the following).

***Measurements***

The critical coding category for the content analysis were 1) frames, analyzed via a qualitative analysis, and 2) emotions in each comment, following the best practices of the previous studies (Chen, Su & Chen, 2019; Lacy et al., 2015; Zhang, Nekmat & Chen, 2020). For the qualitative frame analysis, a pilot content analysis using semi-open coding (Van Gorp, 2010) was performed to define the frame categories and the corresponding emotions.

**Issue-Based Frames**: Preliminary categories identified in a literature review on anti-immigration discourse were applied for the first round of pilot coding. One author of this study and three Singaporeans student coders analyzed 300 comments selected via a stratified and random sampling strategy. During the second phase of pilot coding, the four coders followed the open codification approach (Corbin & Strauss, 2008), repeatedly read and reread the comments, and had several rounds of back-and-forth discussions. Frame categories were merged and created until coders agreed that the results were saturated.

Eventually, we identified the following four frame categories often mentioned online when Singaporeans discussed migrants (See Table 2): (1) *attitudes towards authorities*, including comments that express opinions of how Singapore's authorities work for the people as well as the authorities' abilities and fitness to run, govern, manage Singaporeans issues (2) *attitude towards immigrants*, commenters specifically talk about their attitudes towards immigrants, such as judgment about positive or negative contributions of migrants in Singapore; (3) *co-opted marginality*, comments focus on sentiments expressing injustice compared to immigrants who take advantage of Singapore and/or mentions of government being more lenient or helpful towards immigrants, (4) *identification*, commenters make identity statements about the boundaries and values of the Singaporean identity, such as (a) the use of collective pronouns such as "we," "us" and "our" or (b) value statements about Singaporeans or life in Singapore, including favorability of beliefs about own social grouping or patriotism, (c) depiction of group's interactional dynamics, (e.g. conflicts around race, show of solidarity) (d) the tone of the comments (e.g. friendly, rude) when discussing and negotiating issues with fellow Singaporeans online.

**Emotions**: The valence of each above mentioned 300 comments were classified into positive and negative and determined by our trained coders according to the following principles (See Table 2): (1) *attitude towards immigrant s positive,* if comments emphasize a positive attitude towards/from immigrants and openness towards migrants, including the show of appreciation or support, empathy, sympathy towards migrants and immigrants. *Negative,* if the comment expresses xenophobic ideas / dismiss other races/cultures/nationalities, which is expressed via strong discriminatory language and displays of generalized prejudice; (2) *attitudes towards authorities positive*, if authorities are depicted as helpful, good, sympathetic, competent, or if the comments recognize the importance and the quality of the work accomplished by authorities and institutions in Singapore. *Negative,* if the comment mentions the government/institutions to be self-serving and not considering the interests of Singaporeans, or if authorities are criticized for lack of competency; (3) *identification*, *positive,* if the comment is showing patriotism or talk about positive experiences about Singaporean citizens, like (a) appreciation/empathy for other Singaporeans (work of teachers, the plight of older people), (b) pride and patriotic feelings towards Singapore, open positive affection towards Singapore (c) supportive, sympathetic and friendly interactions with other group members. *Negative,* if the commenters dismiss or criticize Singaporeans/Singapore's, such as (a) judging negatively Singapore quality of life (b) dismiss and criticize other Singaporean citizens (bad attitudes, emphasized racial or any other divisions).

**Table 2**

Examples and Keywords of Each Frame

| Frames | Representative Keywords | Example of Comments: (1) positive (2) negative |
|---|---|---|
| Attitudes Towards Authorities | *PAP, work hard, accomplishment, good for the country, high salary, take care, competent, self-interested, authoritarian* | (1) *Singaporeans are very blessed, a kind hearted government work very hard to take care their people, care every single things to make the better life;* <br> *(2) So, they are telling us to live within our means and tighten our belts, as they continue to raise the COST of living and increase prices of essentials… While insinuating that they should be paid a much more higher SALARY, despite being the most highest paid government leaders in the world. Go figure.* |
| Attitude Towards Immigrants | *contribution, jobs, employment, wages, economy, manners, overcrowding, respect, loud, it used to be better, security, public transport, helpful, multicultural, birth rate* | (1) *Who says china people are barbaric? i personally work with some of them here and most are helpful and polite even the cleaners at my work place gives me a smile when we walk pass each other;* <br> *(2) How to ensure a higher chance for food poisoning… Hire foreign cheap poorly trained staffs who does not understand the importance of hygiene, slaves the staffs into long hours of WORK, source food from the cheapest and last but not least get it delivered in the cheapest possible way on vehicles not designed for food delivery.* |
| Co-opted Marginality | *second class citizens, foreign talents,* | *Exactly… And JOBS are readily for immigrants first… Native Singaporeans get discrimination. What a country. Immigrants, PR'S and New Citizens help to hire more immigrants. Our government got it wrong or purposely overlook the fact that it's other "FOREIGN talents" that are taking our executive positions instead.* <br> *This is wat foreign ft do best. just come, sg is lovely n home to them. make money liao, home* |

| | | *become China n India. always giving these ft money making chances instead of giving sg ppl. sigh…* |
| Identification | *we, us, Singaporeans, proud, pPatriotic, singapore, heartland, achievement, kpo, depreciation, country, thank you, Singapore, together, 70%, sacrifice, race* | *(1) This is just a small investment towards world peace. I am very proud for a tiny Singapore to make a significant CONTRIBUTION to world peace* *(2) A national costume should have something which stands out as an icon for the country. If you put the merlion, that is an icon. You put two flags of FOREIGN countries, does that even have any Singaporean identity?* |

The coding scheme was iteratively developed and pilot-tested with 125 randomly sampled comments from the final dataset. One author of this study and three Singaporean student coders were trained and coded the comments independently. The last round included 125 comments with satisfactory Krippendorff alpha inter-rater reliability test among the four coders (Neuendorf, 2002): Attitudes Towards Authorities (Authorities Negative, α =.93; Positive, α =.87), Attitude Towards Immigrants (Negative, α =.85; Positive, α =.77), Co-opted Marginality (α =.93), Identification (Negative, α =.85, Positive, α =.77)

Additionally, we randomly sampled 15,596 comments from the Entire Corpus. Four human judges coded separately whether a comment belonged to Attitude Toward Authorities, Attitude Toward Immigrants, Co-opted marginality, and Identification (i.e., if a comment belongs to Identification, then coded 1, otherwise 0). Besides, the coders judged each comment's attitude, with positive attitude=1 and negative attitude=0 (about the distribution of the annotated corpus, See Note 1).

**Media Types**: three media types are proposed for our research. We grouped them according to the government's level of financial independence and their expressed value orientations toward the Singapore government: pro-government, neutral, and critical sites.

**Table 3**

Classification of the Sites

| Media Types | Definition | Representative Sites |
| --- | --- | --- |
| Pro-Government (1) | Government-owned media companies | The Straits Time, Channel News Asia, Stomp |

| Neutral ((2) | Independent media companies | Mothership, All Stuff Singapore , MustShareNews , SMRT Feedback by The Vigilanteh, Reddit(Singapore), Independent .sg |
|---|---|---|
| Critical (3) | Alternative media | The Online Citizen, The State Times Review (STR) |

**Computer-Assisted Content Analysis: Based on Supervised Machine Learning**

A computer-assisted content (CAC) analysis and supervised machine learning techniques were conducted to classify the remaining comments within the frame mentioned above categories and expressed emotions. We employed a new semi-supervised approach using Google's Bidirectional Encoder Representations from Transformers (BERT) algorithm, drawn from a widely cited study in machine learning (Devlin et al., 2019). BERT is a pre-trained algorithm that leverages an extensive corpus of English-language (Wikipedia and BooksCorpus) to improve the accuracy of various NLP tasks. BERT has achieved state-of-the-art results on eleven NLP tasks and can be easily fine-tuned to create models for a wide range of NLP tasks (Devlin et al., 2019). Trained models on BERT have been shown to perform well on Tweets, which are short and are likely to contain misspellings, sarcasm, textisms, and slang terms (Gondane et al., 2019). BERT has been rarely used in the classification of social media messages on specific political issues, but few attempts have shown promising results (Gupta et al., 2020)

After iteratively adjusting the parameters, we reached a satisfactory performance of prediction models for each frame. Additionally, BERT was also adopted to train the prediction models for frames and emotions to identify each comment's attitude based on the above-mentioned human-annotated dataset (the model's performance shown in Table 3).

**Table 3**

Performance of the Prediction Classifier for Each Frame and Attitude

|  | Precision | Accuracy | Recall Rate | F1 Score |
|---|---|---|---|---|
| Attitude Toward Authorities | 0.780 | 0.803 | 0.778 | 0.790 |
| Attitude Toward Immigrants | 0.815 | 0.835 | 0.731 | 0.817 |
| Co-opted Marginality | 0.835 | 0.810 | 0.751 | 0.805 |
| Identification | 0.801 | 0.765 | 0.796 | 0.764 |
| Emotion | 0.901 | 0.890 | 0.877 | 0.893 |

## Findings

For RQ1, we examined four salience frames to analyze the rhetoric that appears more salient when comments sections participants talk about migrants. We focused our study on comments which explicitly mentioned immigrants, defined by a list of keywords including commonly used in the local jargon Singlish to describe them (See Note 2, e.g., "migrant," "expat," "Chinese," "Singapore," "PRC," "Ang Mo," "Pinoy,"'FDW," etc.). We filtered in immigrant-related comments from the Entire Corpus, with 57,020 comments retained. After the duplication and removing the null comments, 32,233 comments were retained for further analysis (named The Immigrants Mentioned Corpus in the following).

We then outlined the distribution of comments related to each frame. According to Figure 1, within the comments about immigrants (N=32,233) and among the four types of frames (attitude toward authorities, attitude towards immigrant, co-opted marginality, and identification), identification (N=269,804, 73.90%) appears to be the dominating frame and relatively attracting more attention, followed by the attitude towards immigrant (80,315, 37.72%), attitude toward authorities (67,133, 28.98%), and co-opted marginality (44,489, 28.13%). We applied the same analysis to the entire corpus (386,922 comments).

Secondly, we calculated each comment's frame attitude and compared comments to delineate the different emotions among each frame. The results show that compared to attitude toward authorities and co-opted marginality, identification and attitude towards immigrant tends to be slightly more positively framed.

**Figure 1**

Proportion of Each Frame in the Corpus

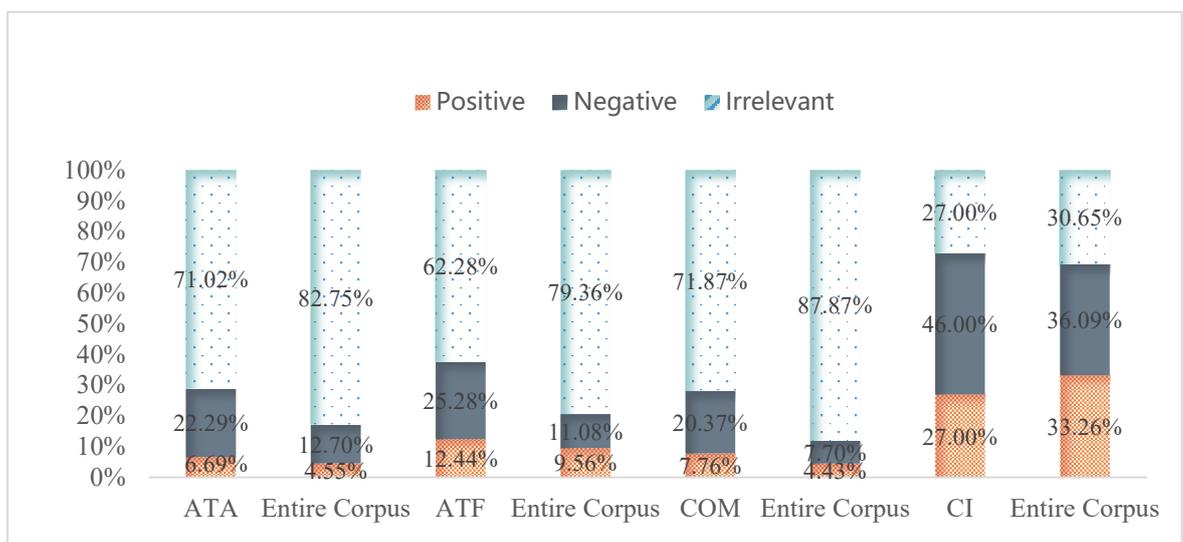

Note. ATA = attitude toward authorities, ATI = attitude towards immigrant, COM= co-opted marginality, and I = identification. Irrelevant indicates the comments that do not belong to a specific frame though belong to an emotion type.

For RQ2, we calculated the mean comment-like count for each frame with different emotions. As Table 2 shows, in general, compared with the comments in the entire corpus, the co-opted marginality frame received noticeably more likes than any other frames in the corpus, implying a relatively higher salience of the discourse.

Moreover, discourse about authorities and identification seems to elicit marginally more reactions when positively rather than negatively framed. For Attitude Towards Immigrants, the situation is the opposite.

**Table 2**

Mean Comment-like Count of Different Types of Discourse and Emotions

| Issue-based Frames | Mean Comment-like Number (SD) | |
| --- | --- | --- |
| | Immigrants Mentioned Corpus | Entire Corpus |
| Attitude Towards Authorities | 104.680 (518.300) | 68.132 (361.201) |
| Positive | 104.790 (577.104) | 85.550 (393.916) |
| Negative | 93.838 (498.826) | 61.887 (348.517) |
| Attitude Towards Immigrant | 128.385 (625.602) | 100.10 (473.864) |
| Positive | 122.185 (550.188) | 93.231 (434.754) |
| Negative | 131.436 (659.568) | 106.021 (505.109) |
| Co-opted Marginality | 149.532 (763.544) | 97.821 (505.691) |
| Positive | 175.258 (732.101) | 94.308 (476.826) |
| Negative | 139.734 (738.049) | 100.045 (523.139) |
| Identification | 146.828 (682.095) | 92.800 (456.934) |
| Positive | 154.953 (681.803) | 85.815 (424.216) |
| Negative | 142.094 (682.243) | 99.231 (485.049) |
| N | 32,233 | 389,031 |

In response to RQ3, Chi-Square tests were conducted to distinguish the differences between the frames by site typology. Results from Table 5 show that attitudes towards authorities frame were more likely to be mentioned on critical sites (31.66%, χ2 = 17426.89, p < 0.001) like *The Online Citizen* and *The State Times Review,* followed by the Neutral sites (13.15%) like *Mothership* and *All Stuff Singapore*, and Pro-Government sites (12.23%) like *The Straits Time, Channel News Asia.* Similarly, co-opted marginality is marginally more discussed by the critical sites users (*14.26%, χ2 = 17426.89, p < 0.001*). In comparison, Attitude Towards Immigrants seems to appear marginally in equal measures in the three types of site such as in Neutral sites (*22.03%, χ2 = 464.242, p < 0.001*), followed by the Pro-Government sites (*19.98%, χ2 = 464.242, p < 0.001*) and Critical sites (18.67%, *χ2 = 464.242, p < 0.001*). Identification seems also to be mentioned in equal measure regardless of the orientation of the site (67,21 %, *70.53%, 69,25%, χ2 =* 366.217, *p < 0.001*).

The Chi-Square tests discriminate against the differences of positive frames usage by site type. As expected, the test shows, in comparison, that critical sites were more frequently negative when discussing attitude toward authorities (*25.16%, χ2 = 1063.783, p < 0.001*), followed by the Neutral (9.39%, *χ2 = 1063.783, p < 0.001*) and Pro-Government sites (8.01%, *χ2 = 1063.783, p < 0.001*).

**Table 5**

χ2 Test of Differences between Frames, Frame's Negativity by Site Type

| Frame Type | Site Type (N=38714) | | | χ2 |
| --- | --- | --- | --- | --- |
| | Pro-Government (N=112976) | Neutral (N=184409) | Critical (N=91329) | DF=2 |
| Attitude Towards Authorities | 12.23% (13,854) | 13.15% (24,258) | 31.66% (28,915) | 17426.898*** |
| Attitude Towards Immigrants | 19.98% (22,568) | 22.03% (40,625) | 18.67% (17,051) | 464.242*** |
| Co-opted Marginality | 9.78% (11,053) | 11.04% (20,361) | 14.26% (13,024) | 1052.694*** |
| Identification | 67.21% (75937) | 70.53% (130057) | 69.65% (63611) | 366.217*** |
| Negative Attitude Towards Authorities | 8.01% (9,047) | 9.39% (1,7312) | 25.16% (22,975) | 1063.783*** |
| Negative Attitude Towards Immigrants | 9.65% (10,899) | 11.01% (20,310) | 12.98% (11,853) | 2204.689*** |
| Negative Co-opted Marginality | 5.12% (5,785) | 6.25% (11,527) | 10.83% (9,894) | 1742.377*** |
| Negative Identification | 30.31% (34,244) | 34.20% (63,081) | 47.02% (42,945) | 8220.317*** |

Note: * p < 0.05, ** p < 0.01, *** p < 0.001. Sample sizes appear in parenthesis below frame percentages.

Results from χ2 tests (Table 6) show that comments on attitude toward immigrants are marginally more likely to be associated with negative identification (23.20%, χ2=1543.52, p < 0.001) than positive identification (21.22%, χ2=1543.52, p < 0.001). Comments on attitude toward authorities are more likely to be associated with negative identification (23.56%, χ2=1035.433, p < 0.001) than positive identification (8.96%, χ2=1035.433, p < 0.001), and co-opted marginality comments tends more likely associated to negative identification (14.64% χ2=2407.733, p < 0.001) than positive identification (10.44%, χ2=2407.733, p < 0.001). Hence, H4b, and H4c are supported, and H4a is rejected.

**Table 6**

Chi-Square Test of Differences between Attitude Towards Authorities, Attitude Towards Immigrants, and Co-opted Marginality by Identification

| Frames | Identification | | | χ2 |
|---|---|---|---|---|
| | Positive(N=129399) | Negative (N=140405) | Irrelevant(N=119227) | DF=2 |
| Attitude Towards Authorities | 8.96% (N=11599) | 23.56% (N=33092) | 16.00% (N=22422) | 10353.433*** |
| Attitude Towards Immigrants | 21.22% (N=27456) | 23.20% (N=32572) | 14.45% (N=20287) | 1543.527*** |
| Co-opted Marginality | 10.44% (N=13513) | 14.64% (N=20552) | 7.42% (N=10418) | 2407.773*** |

## Discussion and Conclusion

This study contributes to the literature on immigrants' discourse on social media by suggesting an analytic framework to grasp its complexity and nuances. The four frames of co-opted marginality, attitude towards immigrants, attitude towards authorities, and identification allow synthesizing conceptual blocks usually isolated when studying migrants' attitudes. As mentioned earlier, the study of anti-immigration discourse on social media has tended to emphasize different facets of the issues attached to anti-immigration sentiments such as populism, authoritarian attitudes, prejudices towards races, or extremist political activities.

Second, we focused on grasping the complexity and the multiple facets of nativist discourse on social media. Based on qualitative and quantitative frame analysis, our empirical analysis of the public discourse in online public spheres confirmed the co-presence of many frames when Singaporeans talk about immigrants online. The computer-assisted content analysis indicates that with six months' worth of comments across all types of platforms, 29.28% of the corpus linked to immigrants show negative discourse about immigrants; 20.37% is attributable to co-opted marginality (feelings of inferiority towards immigrants); 22.29 % to a negative attitude towards authorities and 46.69% to negative collective identification. The co-opted marginality frame's importance may call for more examination on the nature of the anti-immigrant discourse rather than condemnation.

Although largely ignored by the dominant discourse, the co-opted marginality frame is widely used when Singaporeans comment online. Claims of marginalization by dominant ingroup members still need to be further scrutinized. We can only postulate whether it is a mild, disguised collective expression of anti-immigrant sentiment made palatable and acceptable by the dominant consensus. As mentioned earlier, racist, and anti-immigrant social media discourse gets increasingly sophisticated (Bliuc et al., 2018). According to the Simon Wiesenthal Center's Digital Hate and Terrorism (2018), Younger generations are restrained in the use of hateful language; they resort to micro-aggressions that are almost undetectable (Breitfeller et al., 2019) but present high toxicity and potency in denigrating minorities (Clark et al., 2011).

The co-opted marginality discourse is visible across platforms where commenting activity occurs. This emphasizes the importance of the diffusion and the ubiquity of the frame in the online public sphere. Discourse about migrants in comment sections seems to be homogeneously distributed regardless of the community sites' leanings (pro-government, independent, or critical). The discourse's omnipresence suggests the frames' salience in the public sphere and may reveal a widely shared collective cognitive structure.

Finally, we began this article by developing a theoretical framework arguing that comment sections can constitute a subaltern space for citizens' public and political engagement. Our study is based on data collected in Singapore, whose media environment is controlled by the government. Since the city-state independence, the PAP (People's Action Party), Singapore's main party in power, has become the single most powerful influence on the evolution of the public sphere by "molding a sophisticated press control regime" (Tey, 2008, p876) to push for a pragmatic political ideology. The effort has shaped a compliant public sphere whereby journalists do not see political advocacy and dissent as part of their agenda (Hao & George, 2012). Mainstream media in Singapore claim that Singapore's long-term interests ultimately guide them; there is no shame in supporting a good government to contain threats to the nation-building project (George & Venkiteswaran, 2019). Although our study shows that polity is discussed in the Singaporean comment section and, at times, under a negative and critical light, notably on sites that have asserted a readiness to be independent of the political system (up to 25% of all comments). It is overall less noticeable in the mainstream digital platforms (close to 10% of all comments). The comment section therefore does not seem to clearly constitute a quasi-public sphere for a counter-public, as previously documented in a western context (Ruiz et al., 2011; Teopfl & Piwoni, 2018).

**Limitations**

The study aims to capture the nature of social media's discourse from a constructivist perspective. We do not address questions related to its representativity in society. Whether the comment sections mirror public opinion and match the distribution and the nature of the ideas circulating in the public sphere is an entirely different issue we cannot currently address in our analysis. Recent research points out that user-generated comments significantly impact readers' perceptions of public opinion and their personal opinions (Lee, 2012; Lee & Jang, 2010). As the mainstream media systematically discriminate against the views of minorities (particularly in the context of our study), we argue that comment sections may offer a legitimate means to challenge unjust participatory privileges (Dalberg, 2011; Downey & Fenton, 2003).

Additionally, we do not look at the collateral effects of the platform's architecture nor the gatekeeping effect. The affordances of the online public forums may indeed have a role in the emergence of counter publics. We acknowledge that signaling functions such as "most popular first" or "newest first" may affect how commenters and readers coordinate the flow of information and therefore impact the salience of the issue discussed. This may be an avenue for further research in the specific context of Singapore's online public sphere. Similarly, we did not consider the effect of gatekeepers and how the journalistic or news site articles' content influenced the nature of the counter-publics that emerged in the thread below the articles. As mentioned earlier, our findings suggest negligible differences between the site's political allegiance in the reactions to posts about immigrants and immigrants. Still, we only considered the thread's comments' valence and not one of the original posts.

In conclusion, the paper's findings suggest that discourse about immigrants online may be multifaceted and revelatory of a broader critical discourse within a social context. The notion of immigrant appears to be both a salient object of critique and a collectively shared subject to convey criticisms about the surrounding polity. Our analytical framework can grasp nuances about emerging discourses about immigrants.

## Notes

**1.** About the distribution of annotated samples: (1) Frames: Attitude Towards Authorities(N=2057), Attitude Towards Immigrants(N=1751), Co-opted Marginality(N=1200), Identification(N=9948); (2)Emotions: Positive(N=8979), Negative(N=6617).

2. Local jargon Singlish keywords used to describe immigrants: Ang_moh caucasian ceca china chinese citizen citizens community countries country culture english european expats foreign domestic workers, filipino foreign immigrant helper homeland immigrants india indians indonesia job local maid malaysia malaysian migrants permanent permit philippines pinoy PR PRC race racism racist racists Raper Resident Singaporeans Sinkies talents thambis tiong unemployment visa worker work pass xenophobia.

# References

Alduy, C. (2016). Nouveau discours, nouveaux succès. *Pouvoirs*, (2), 17-29.

Almgren, S. M., & Olsson, T. (2016). Commenting, sharing and tweeting news. *Nordicom Review, 37*(2), 67-81.

Ashmore, R. D., Deaux, K., & McLaughlin-Volpe, T. (2004). An organizing framework for collective identity: articulation and significance of multidimensionality. *Psychological bulletin, 130*(1), 80.

Atton, C. (2006). Far-right media on the Internet: culture, discourse and power. *New media & society, 8*(4), 573-587.

Bartlett, J., Birdwell, J., & Littler, M. (2011). *The new face of digital populism*: Demos.

Bennett, W. L., Segerberg, A., & Walker, S. (2014). Organization in the crowd: peer production in large-scale networked protests. *Information, Communication & Society, 17*(2), 232-260.

Betz, H. G. (2017). Nativism across time and space. *Swiss Political Science Review, 23*(4), 335-353.

Bliuc, A.-M., Faulkner, N., Jakubowicz, A., & McGarty, C. (2018). Online networks of racial hate: A systematic review of 10 years of research on cyber-racism. *Computers in Human Behavior, 87*, 75-86.

Bode, L. (2012). Facebooking It to the Polls: A Study in Online Social Networking and Political Behavior. *Journal of Information Technology & Politics, 9*(4), 352-369.

Boltanski, L. (2011). *On critique: A sociology of emancipation*: Polity.

Boltanski, L., & Claverie, É. (2007). Du monde social en tant que scène d'un procès. In.

Boomgaarden, H. G., & Vliegenthart, R. (2007). Explaining the rise of anti-immigrant parties: The role of news media content. *Electoral studies, 26*(2), 404-417.

Breitfeller, L., Ahn, E., Jurgens, D., & Tsvetkov, Y. (2019). *Finding microaggressions in the wild: A case for locating elusive phenomena in social media posts.* Paper presented at the Proceedings of the 2019 Conference on Empirical Methods in Natural Language Processing and the 9th International Joint Conference on Natural Language Processing (EMNLP-IJCNLP).

Caiani, M., & Parenti, L. (2016). *European and American extreme right groups and the Internet*: Routledge.

Chen, Z., Su, C. C., & Chen, A. (2019). Top-down or bottom-up? A Network agenda-setting study of Chinese nationalism on social media. Journal of Broadcasting & Electronic Media, 63(3), 512-533.

Chin, D. (2017, January 15th). This is Singapore' is no excuse for racist behaviour. *The Straits Times*, https://www.straitstimes.com/singapore/this-is-singapore-is-no-excuse-for-racist-behaviour

Chwe, M. S.-Y. (2001). Rational ritual: culture. *Coordination, and Common*.

Clark, D. A., Spanierman, L. B., Reed, T. D., Soble, J. R., & Cabana, S. (2011). Documenting Weblog expressions of racial microaggressions that target American Indians. *Journal of Diversity in Higher Education, 4*(1), 39.

Cresswell, C., Whitehead, K. A., & Durrheim, K. (2014). The anatomy of "race trouble" in online interactions. *Ethnic and racial studies, 37*(14), 2512-2528.

Dahlberg, L. (2011). Re-constructing digital democracy: An outline of four "positions". *New media & society, 13*(6), 855-872.

Dahlgren, P. (2005). The Internet, Public Spheres, and Political Communication: Dispersion and Deliberation. *Political Communication, 22*(2), 147-162.

Daniels, J. (2013). Race and racism in Internet Studies: A review and critique. *New media & society, 15*(5), 695-719.

Domingo, D., Quandt, T., Heinonen, A., Paulussen, S., Singer, J. B., & Vujnovic, M. (2008). Participatory journalism practices in the media and beyond: An international comparative study of initiatives in online newspapers. *Journalism Practice*, *2*(3), 326-342.

Downey, J., & Fenton, N. (2003). New media, counterpublicity and the public sphere. *New media & society*, *5*(2), 185-202.

Durrheim, K., Greener, R., & Whitehead, K. A. (2015). Race trouble: Attending to race and racism in online interaction. *British Journal of Social Psychology, 54*(1), 84-99.

Earle, M., & Hodson, G. (2020). Questioning white losses and anti-white discrimination in the United States. *Nature Human Behaviour, 4*(2), 160-168.

Engesser, S., Ernst, N., Esser, F., & Büchel, F. (2017). Populism and social media: how politicians spread a fragmented ideology. *Information, Communication & Society, 20*(8), 1109-1126.

Erjavec, K., & Kovačič, M. P. (2012). "You Don't Understand, This is a New War!" Analysis of Hate Speech in News Web Sites' Comments. *Mass Communication and Society, 15*(6), 899-920.

Ernst, N., Blassnig, S., Engesser, S., Büchel, F., & Esser, F. (2019). Populists Prefer Social Media Over Talk Shows: An Analysis of Populist Messages and Stylistic Elements Across Six Countries.

Ernst, N., Engesser, S., Büchel, F., Blassnig, S., & Esser, F. (2017). Extreme parties and populism: an analysis of Facebook and Twitter across six countries. *Information, Communication & Society, 20*(9), 1347-1364.

Ferrari, E. (2016). Social media for the 99%? Rethinking social movements' identity and strategy in the corporate web 2.0. *Communication and the Public, 1*(2), 143-158.

Flesher Fominaya, C. (2010). Collective Identity in Social Movements: Central Concepts and Debates. *Sociology Compass, 4*(6), 393-404.

Freelon, D. (2015). Discourse architecture, ideology, and democratic norms in online political discussion. *New media & society, 17*(5), 772-791.

Fuchs, C. (2018). Authoritarian capitalism, authoritarian movements and authoritarian communication. *Media, culture & society, 40*(5), 779-791.

Gattinara, P. C. (2018). Europeans, shut the borders! Anti-refugee mobilisation in Italy and France. In *Solidarity Mobilizations in the "Refugee Crisis"* (pp. 271-297): Springer.

George, C. (2017). Singapore, incomplete: Reflections on a first world nation's arrested political development.

Giroux, H. A. (2019). Trump and the legacy of a menacing past. *Cultural studies, 33*(4), 711-739.

Goodman, S., & Rowe, L. (2014). "Maybe it is prejudice … but it is NOT racism": Negotiating racism in discussion forums about Gypsies. *Discourse & Society, 25*(1), 32-46.

Gordon, P. E. (2017). The authoritarian personality revisited: Reading Adorno in the age of Trump. *boundary 2, 44*(2), 31-56.

Hallin, D. C., & Mancini, P. (2011). *Comparing media systems beyond the Western world*: Cambridge University Press.

Halpern, D., Valenzuela, S., & Katz, J. E. (2017). We face, I tweet: How different social media influence political participation through collective and internal efficacy. *Journal of Computer-Mediated Communication, 22*(6), 320-336.

Hao, X., & George, C. (2012). Singapore journalism: Buying into a winning formula. In *The global journalist in the 21st century* (pp. 101-113). Routledge.

Herkman, J. P. (2016). Construction of Populism. *Nordicom Review*.

Hogg, M. A. (2016). Social identity theory. In *Understanding peace and conflict through social identity theory* (pp. 3-17). Springer, Cham.

Inglehart, R., & Norris, P. (2016). Trump, Brexit, and the rise of populism: Economic have-nots and cultural backlash.

Kavada, A. (2015). Creating the collective: social media, the Occupy Movement and its constitution as a collective actor. *Information, Communication & Society, 18*(8), 872-886.

Kelly Garrett, R. (2006). Protest in an information society: A review of literature on social movements and new ICTs. *Information, Communication & Society, 9*(02), 202-224.

Kirk, R., & Martin, S. A. (2017). The dark power of words: Stratagems of hate in the 2016 presidential campaign. In *The 2016 US Presidential Campaign* (pp. 205-229). Palgrave Macmillan, Cham.

Klein, O., & Muis, J. (2018). Online discontent: comparing Western European far-right groups on Facebook. *European Societies*, 1-23.

Laaksonen, S.-M., Pantti, M., & Titley, G. (2020). Broadcasting the Movement and Branding Political Microcelebrities: Finnish Anti-Immigration Video Practices on YouTube. *Journal of communication*.

Lacy, S., Watson, B. R., Riffe, D., & Lovejoy, J. (2015). Issues and best practices in content analysis. *Journalism & Mass Communication Quarterly*, *92*(4), 791-811.

Lee, E. J. (2012). That's not the way it is: How user-generated comments on the news affect perceived media bias. *Journal of Computer-Mediated Communication*, *18*(1), 32-45.

Lee, E. J., & Jang, Y. J. (2010). What do others' reactions to news on internet portal sites tell us? Effects of presentation format and readers' need for cognition on reality perception. *Communication research*, *37*(6), 825-846.

Lind, M. (2020). *The New Class War: Saving Democracy from the Managerial Elite*: Penguin.

Linda, Y. (2015). *Singapore's economic development: Retrospection and reflections*: World Scientific.

Martin, J. L. (2001). The Authoritarian Personality, 50 Years Later: What Questions Are There for Political Psychology? *Political Psychology, 22*(1), 1-26.

Mastro, D., & Tukachinsky, R. (2011). The influence of exemplar versus prototype-based media primes on racial/ethnic evaluations. *Journal of communication, 61*(5), 916-937.

Matthes, J., & Schmuck, D. (2017). The Effects of Anti-Immigrant Right-Wing Populist Ads on Implicit and Explicit Attitudes: A Moderated Mediation Model. *Communication research, 44*(4), 556-581.

McNamee, L. G., Peterson, B. L., & Peña, J. (2010). A Call to Educate, Participate, Invoke and Indict: Understanding the Communication of Online Hate Groups. *Communication Monographs, 77*(2), 257-280.

Moussaïd, M., Brighton, H., Gaissmaier, W., 2015. The amplification of risk in experimental diffusion chains. Proceedings of the National Academy of Sciences 112, 5631–5636.. doi:10.1073/pnas.1421883112

Neuendorf, K. A. (2002). Defining content analysis. Content analysis guidebook. Thousand Oaks, CA: Sage.

Noci, J. D., Domingo, D., Masip, P., Micó, J., & Ruiz, C. (2012). *Comments in news, democracy booster or journalistic nightmare: Assessing the quality and dynamics of citizen debates in Catalan online newspapers.* Paper presented at the International Symposium on Online Journalism.

Papacharissi, Z. (2015). *Affective publics: Sentiment, technology, and politics*: Oxford University Press.

Quinsaat, S. (2014). Competing news frames and hegemonic discourses in the construction of contemporary immigration and immigrants in the United States. *Mass Communication and Society*, *17*(4), 573-596. https://doi.org/10.1080/15205436.2013.816742

Reilly, K. (2016, September 10th). "Read Hillary Clinton's 'Basket of Deplorables' Remarks About Donald Trump Supporters," TIME, http://time.com/4486502/hillary-clinton-basket-of-deplorables-transcript/.

Rieder, B. (2013, May). Studying Facebook via data extraction: the Netvizz application. In Proceedings of the 5th annual ACM web science conference, 346-355.

Riles, J. M., Pilny, A., & Tewksbury, D. (2017). Media fragmentation in the context of bounded social networks: How far can it go? *New media & society, 20*(4), 1415-1432.

Rodan, G. (1998). The Internet and political control in Singapore. *Political Science Quarterly, 113*(1), 63-89.

Ruiz, C., Domingo, D., Micó, J. L., Díaz-Noci, J., Meso, K., & Masip, P. (2011). Public sphere 2.0? The democratic qualities of citizen debates in online newspapers. *The international journal of press/politics, 16*(4), 463-487.

Schemer, C. (2012). The Influence of News Media on Stereotypic Attitudes Toward Immigrants in a Political Campaign. *Journal of communication, 62*(5), 739-757.

Scullion, R., Gerodimos, R., Jackson, D., & Lilleker, D. (Eds.). (2013). *The media, political participation and empowerment*. Routledge.

Serwer, A. (2019). White nationalism's deep American Roots. *The Atlantic*.

Spry, D. (2019). More Than Data: Using the Netvizz Facebook Application for Mixed-Methods Analysis of Digital Diplomacy. SAGE Publications Ltd.

Stage, C. (2013). The online crowd: a contradiction in terms? On the potentials of Gustave Le Bon's crowd psychology in an analysis of affective blogging. *Distinktion: Journal of Social Theory, 14*(2), 211-226.

Teng, A. (2019, 21may). The Big Read: The immigrant issue—are we ready for a rethink? *TODAY online*, https://www.todayonline.com/singapore/big-read-immigrant-issue-are-we-ready-rethink

Tey, T. H. (2008). Confining the Freedom of the Press in Singapore: A" Pragmatic" Press for" Nation-Building"?. *Human Rights Quarterly*, 876-905.

Thurman, N. (2008). Forums for citizen journalists? Adoption of user generated content initiatives by online news media. *New media & society, 10*(1), 139-157.

Toepfl, F., & Piwoni, E. (2015). Public spheres in interaction: Comment sections of news websites as counterpublic spaces. *Journal of communication, 65*(3), 465-488.

Toepfl, F., & Piwoni, E. (2018). Targeting dominant publics: How counterpublic commenters align their efforts with mainstream media. *New media & society, 20*(5), 2011-2027.

Treem, J. W., & Leonardi, P. M. (2013). Social media use in organizations: Exploring the affordances of visibility, editability, persistence, and association. *Annals of the International Communication Association, 36*(1), 143-189.

Tynes, B. M., Umana-Taylor, A. J., Rose, C. A., Lin, J., & Anderson, C. J. (2012). Online racial discrimination and the protective function of ethnic identity and self-esteem for African American adolescents. *Developmental psychology, 48*(2), 343.

Vadaketh, S. T., & Low, D. (2014). *Hard choices: Challenging the Singapore consensus*. NUS Press.

Vaswani, K. (2019). Concern over Singapore's anti-fake news law, *BBC News ,* https://www. bbc. com/news/business-47782470

Warner, M. (2002). Publics and counterpublics. *Public culture, 14*(1), 49-90.

Weaver, S. (2013). A rhetorical discourse analysis of online anti-Muslim and anti-Semitic jokes. *Ethnic and racial studies, 36*(3), 483-499.

Whooley, O. (2007). Collective identity. *The Blackwell encyclopedia of sociology*.

Wirz, D. S., Wettstein, M., Schulz, A., Müller, P., Schemer, C., Ernst, N., … Wirth, W. (2018). The Effects of Right-Wing Populist Communication on Emotions and Cognitions toward Immigrants. *The international journal of press/politics, 23*(4), 496-516.

Zhang, X., Nekmat, E., & Chen, A. (2020). Crisis collective memory making on social media: A case study of three Chinese crises on Weibo. Public Relations Review, 46(4), 101960.

Zhou, Y., Reid, E., Qin, J., Chen, H., & Lai, G. (2005). US domestic extremist groups on the Web: link and content analysis. *IEEE Intelligent Systems, 20*(5), 44-51.

Ziegele, M., & Quiring, O. (2013). Conceptualizing online discussion value: A multidimensional framework for analyzing user comments on mass-media websites. *Annals of the International Communication Association, 37*(1), 125-153.